 \documentstyle[aps,prc,twocolumn,floats,epsfig]{revtex}
\begin{document}
%
%==========================================================================
%
\renewcommand{\thefootnote}{\arabic{footnote}}

\twocolumn[\columnwidth\textwidth\csname@twocolumnfalse\endcsname

\title{Pairing Gap and Polarisation Effects}
\author{K. Rutz${}^{1}$,
        M. Bender${}^{2,3}$,
        P.--G. Reinhard${}^{4,5}$,
        J. A. Maruhn${}^{1,5}$
}
\address{${}^{1}$Institut f\"ur Theoretische Physik,
         Universit\"at Frankfurt,
         Robert--Mayer--Str.\ 10, D--60325 Frankfurt am Main, Germany
}
\address{${}^{2}$Department of Physics and Astronomy,
         The University of North Carolina,
         Chapel Hill, NC 27599, U.S.A.
}
\address{${}^{3}$Department of Physics and Astronomy,
         The University of Tennessee,
         Knoxville, TN 37996, U.S.A.
}
\address{${}^{4}$Institut f\"ur Theoretische Physik II,
         Universit\"at Erlangen--N\"urnberg,
         Staudtstr.\ 7, D--91058 Erlangen, Germany
}
\address{${}^{5}$Joint Institute for Heavy--Ion Research,
         Oak Ridge National Laboratory,
         P.\ O.\ Box 2008, Oak Ridge, TN 37831, U.S.A.
}

\date{October 7, 1999}

\maketitle

%\addvspace{2mm}
%
%=======================================================================
%
\begin{abstract}
The phenomenological adjustment of the nuclear pairing strength is
usually performed with respect to the odd-even staggering of the
binding energies. We find that the results strongly depend on the way
in which the ground states of the odd nuclei are computed. A thorough
calculation including all time-even and time-odd polarisation effects
induced by the odd nucleon produces about $30 \%$ reduced odd-even
staggering as compared to the standard spherical calculations in the
relativistic mean-field model. The pairing strength must be enhanced
by about $20 \%$ to compensate for that effect. The enhanced strength
has dramatic consequences for the predicted deformation properties of
the underlying mean-field models, possibly implying that new
adjustments of their parameters become necessary as well.
\end{abstract}
\addvspace{2mm}


\addvspace{5mm}]

\narrowtext
%
%=======================================================================
%

Pairing is an essential ingredient of nuclear mean-field models
since the early days of the single-particle shell model
\cite{Boh58a,Bel58a}. It can be motivated from a theoretical point of
view by a near-singularity of the effective $T$ matrix near the Fermi
surface \cite{Thouless} and it manages to subsummarize a large amount
of two-body correlations at the price of a moderately extended
mean-field theory. From a phenomenological point of view, pairing is
needed to explain, e.g., the existence of spherical non-magic nuclei
and to account for the observed strong odd-even staggering of the
binding energies. The purely theoretical access to pairing from
nuclear many-body theory is still plagued by unresolved quantitative
problems of nuclear ``ab initio'' calculations \cite{revcorr}. One
thus has to recur to phenomenological information which is exploited
nowadays with a high degree of systematics (for a recent and
comprehensive compilation see \cite{MadNix}). The general strategy of
these phenomenological evaluations is to relate the odd-even
staggering of the binding energies to the pairing gap by taking
appropriate differences, e.g., for neutrons the fourth order
difference
\begin{eqnarray}
\label{eq:gap}
\Delta^{(4)}_n
& = & - {\textstyle \frac{1}{8}} (-1)^N
	\big[ E(N-2)-4E(N-1)+6E(N)
      \nonumber\\
&   & \qquad\qquad\qquad
      - 4 E(N+1) + E(N+2) \big]
\end{eqnarray}
at fixed proton number. This access, however, has the basic problem
that pairing is not the only source of odd-even staggering. Odd-even
fluctuations in energy can also be produced by polarisation of the
even core through the odd nucleon, which leads to odd-even jumps in
deformation, spin-alignment or dynamical currents. For example, there
is pronounced odd-even staggering in the energies of metal clusters
which, however, can be explained exclusively by effects of
spin-alignment and Jahn-Teller deformation \cite{KohRei}. Spin
alignment can be excluded for most nuclei.  The Jahn-Teller effect
alone can produce sizeable odd-even effects in deformed nuclei as was
worked out in a recent paper \cite{Dobac}, but its influence is
diminished for semi-magic nuclei if pairing is switched on because
this restores spherical symmetry over the whole isotopic or isotonic
chain. Nonetheless, there remains a polarisation of the core through
the field of the odd nucleon, which may deliver substantial
contributions to the odd-even staggering and thus should be taken into
account when adjusting pairing strengths to the phenomenological
pairing gap (\ref{eq:gap}). The question is how large these
polarisation effects are in practice. There is, first, the static
deformation-polarisation which is induced by the finite multipole
moment of the odd nucleon's density.  In addition the odd nucleon
breaks the intrinsic time-reversal invariance because it carries a
nonzero spin and a current contribution. These time-odd components can
induce a sizeable time-odd response in terms of spin and current
polarisation in the even-even core \cite{ugpap}. In the following we
call that a dynamical polarisation to distinguish it from the mere
deformation effects. It is the aim of this letter to investigate the
effect of such polarisation effects on the odd-even staggering and
thus on the adjustment of the pairing gap (\ref{eq:gap}).

The investigation requires a mean-field model which provides reliable
response properties in all channels, including unnatural parity states
and spin polarisation. The relativistic mean-field (RMF) model includes a
``natural'' description of spin properties \cite{ugpap,Ringspin,Afa98a} 
and is a reliable starting point for our present investigation of dynamical
polarisation. A similar investigation on the grounds of the 
non-relativistic Skyrme-Hartree-Fock model was performed 
recently \cite{Xu99a}.
We employ the RMF with the parametrization {PL-40}
\cite{PLpap} similar as in a previous systematic investigation of odd
nuclei near magic shells \cite{ugpap}. The details of the model, the
strategy to define the blocking in odd nuclei, and the numerical
handling can be found there. We perform axially symmetric deformed
calculations including time-odd currents to take into account the
broken time-reversal symmetry and its related polarisation effects.
Pairing is described at the BCS level using matrix elements computed
from a zero-range pairing force ${V_q \, \delta({\bf r}_1-{\bf r}_2)}$
with ${q\in\{p,n\}}$ acting in a pairing space which is limited by a
soft cut-off of Woods-Saxon shape in energy \cite{Ton79a,Kri90a} with
the cut-off adjusted dynamically to include ${N_q + 1.65 \,N_q^{2/3}}$
nucleon states \cite{cutoff}. The pairing strengths are fitted to the
experimental gaps within several semi-magic isotope and isotone chains
in connection with the actual mean-field force. The odd nuclei needed
for this approach are calculated in a spherical approximation.  For
{PL-40}, which is used in this study, we find ${V_p = -348\ {\rm
MeV\,fm}^3}$ and ${V_n = -346\ {\rm MeV\, fm}^3}$. These values
correspond to a standard determination of pairing strengths without
polarisation effects. We take them as a base point for comparison with
the more elaborate adjustments to be discussed in the following.
\begin{table}[t!]
\begin{tabular}{lcc}
 approach & $V_n$ $({\rm MeV}\,{\rm fm}^3)$ & $\Delta^{(4)}_n$ (MeV)\\
\hline
 exp.       &           & 1.2587 \\[0.3ex]
 spherical   & $-346.0$  & 1.2493 \\
 deformed    & $-346.0$  & 1.1600 \\
             & $-346.0$  & 0.8104 \\
 \raisebox{1.5ex}[-1.5ex]{dyn. polarisation} & $-412.4$  & 1.2588
\end{tabular}
\caption{\label{tab:refit}
Neutron gaps $\Delta^{(4)}_n$ in $^{126}{\rm Sn}$ for the various
levels of approximation. In case of dynamical polarisation the gaps
are shown for the original as well as for the refitted pairing
strength $V_n$ (last line).}
\end{table}

As a test case we consider the neutron gaps in the chain of Sn
isotopes. The magic proton number in Sn allows to concentrate on the
neutron gaps and it renders all even isotopes spherical, which
confines the deformation effects exclusively to the odd isotopes.

\begin{figure}[t!]
\centerline{\epsfig{figure=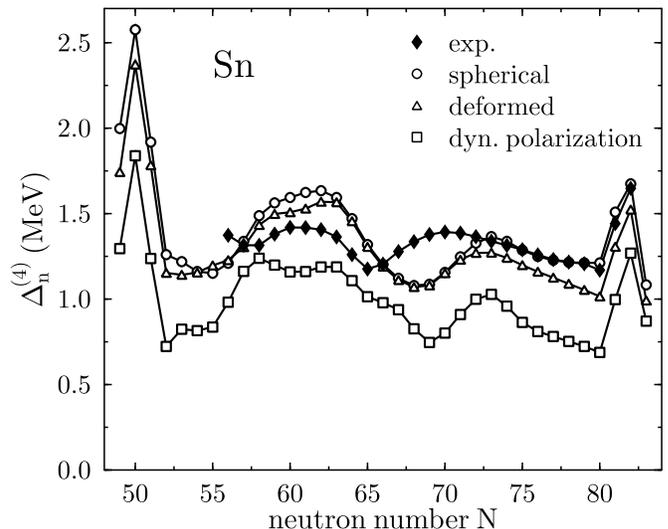}}
\caption{\label{fig:gapn}Fourth-difference neutron gaps
(\protect\ref{eq:gap}) for Sn isotopes
(${Z=50}$) computed with the RMF parametrization \mbox{PL-40} 
in different approaches.  Spheres stand for purely spherical calculations
without time-odd currents, triangles for deformed calculations
without time-odd currents, and squares for unrestricted deformed
calculations including time-odd currents.  Experimental gaps are
given as full rhombi for comparison (the masses were taken from
\protect\cite{masstab}).}
\end{figure}

Figure \ref{fig:gapn} shows the results for the neutron gap
(\ref{eq:gap}) at various levels of mean-field calculations:
spherically, deformed but time-even, and deformed with dynamical
polarisation (i.e., including time-odd currents). As mentioned above,
the spherical results had been fitted to agree with the experimental
gaps in the average, which has been achieved more or less
nicely. Allowing for deformation (yet without dynamical polarisation)
indeed changes the gaps at several isotopes.  For the heavier nuclei,
however, to which pairing is usually fitted, the resulting reduction
of the gap is sufficiently small to neglect them at the level of
quality with which pairing can be adjusted anyway.  This can be
deduced from the small changes from ``spherical'' to ``deformed'' seen
in fig.~\ref{fig:gapn}. The effect is much larger when allowing for
dynamical polarisation.  The gaps shrink in average by about $30\%$.
The example here shows that dynamical polarisation can have a large
effect on the odd-even staggering which, in turn, puts a warning signs
on a phenomenological adjustment of pairing properties on the the
basis of energy differences. One should, in fact, adjust the gaps
(\ref{eq:gap}) anew while using fully polarised calculations.

In order to check that this reduction of the gap is not a particular
feature of {PL-40}, we have performed similar calculations for the
parametrization NL3 \cite{NL3} which was adjusted with different 
bias. we find for the case of ${}^{126}$Sn the sequence 1.32 MeV
for the gap from spherical calculations, over 1.30 MeV for the gap
from deformed and time-even calculations down to 0.93 MeV for the 
case with full dynamical polarisation. That is very much the same 
reduction as explored for {PL-40}, see Table~\ref{tab:refit}.
we thus are confident that the effect is of generic nature, at 
least for the RMF.
\begin{figure}[t!]
\centerline{\epsfig{figure=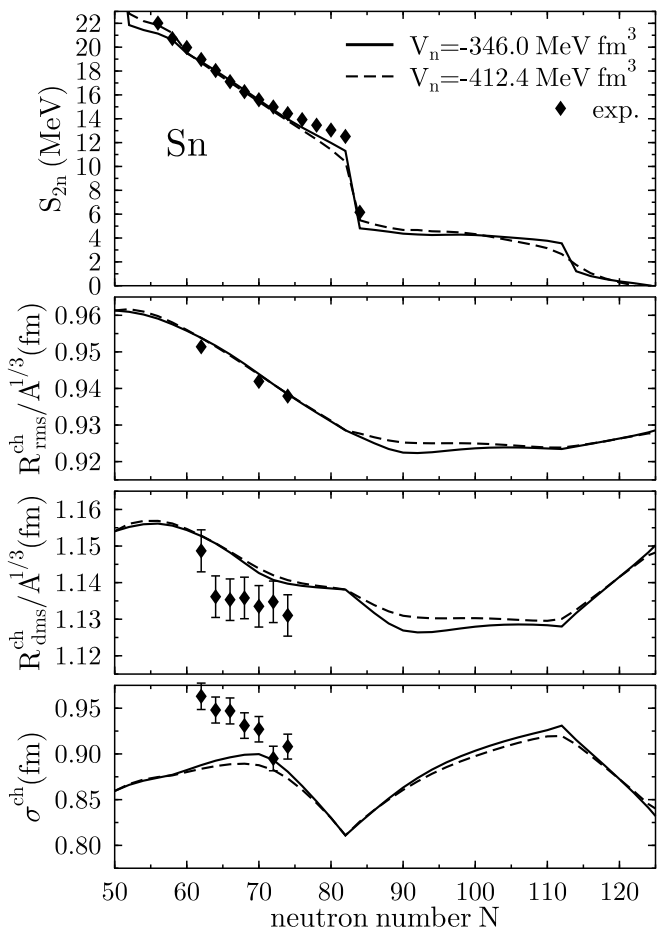}}
\caption{\label{fig:obsrefit}
Ground state observables in Sn isotopes drawn as functions of the
neutron number $N$. Upper panel: two-neutron separation energies
${S_{2n}=E(N-2)-E(N)}$; second from above: rms radii of the charge
distribution, $R_{\rm rms}^{\rm ch}$; third panel: diffraction radii
$R_{\rm dms}^{\rm ch}$; lowest panel: surface thickness $\sigma$.  The
error bars represent the average precision of the mean field model to
reproduce these quantities (the actual experimental errors would be
much smaller).  Two cases are compared: a calculation with the
spherically fitted pairing strength (full lines) and with the refitted
strength (dashed lines).}
\end{figure}

We now want to exemplify the consequences of such a modified pairing
strength for the present test case. To that end, we have performed a
readjustment of the pairing strength taking properly into account all
polarisation effects.  In order to cope with the rather large expense
of the full-fledged calculations of odd nuclei, we have decided to
concentrate on the neutron pairing strength and to refit that with
respect to one nucleus, namely $^{126}{\rm Sn}$. We find that the
readjustment (including time-odd currents) increases the required
pairing-strength parameter by as much as $19\%$.  Proton pairing will
be needed in two applications later on. We find that the same
enhancement factor of $19\%$ reproduces gaps in heavy nuclei very
well.  For that we assume the same upscaling factor as was found for
neutron pairing.
This minimal fitting strategy is sufficient for the present purposes
of an exploratory study.  The results for the case of neutron pairing
in $^{126}{\rm Sn}$ are summarised in table~\ref{tab:refit}.  One sees
that the dynamical polarisation reduces the gap by $35\%$ in this case
and a counteracting readjustment increases the pairing-strength
parameter by $19\%$. It remains to check the consequences for other
observables.
\begin{figure}[t!]
\centerline{\epsfig{figure=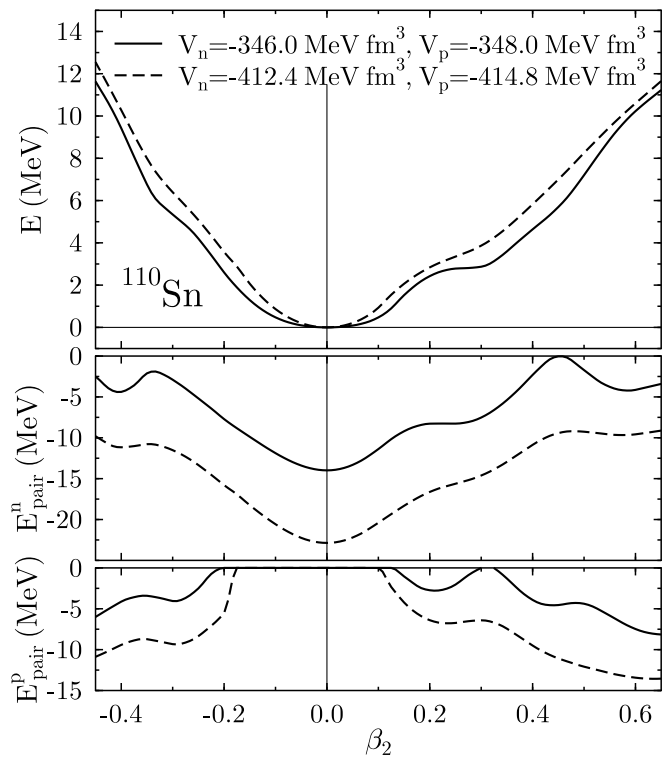}}
\caption{\label{fig:110sn}
Quadrupole deformation properties for $^{110}{\rm Sm}$ drawn versus
the dimensionless quadrupole deformation
${\beta_2=\bigl[4\pi/(3AR^2)\bigr]\langle r^2Y_{20}\rangle}$ with
${R=1.2~{\rm fm}~A^{1/3}}$.  The full line shows the result obtained
with standard pairing (fitted to spherically calculated nuclei only)
and the dashed line those for the enhanced pairing strength (fitted
including polarisation effects).  Upper: Deformation energy, i.e. total
energy rescaled to zero at the spherical minimum.  Middle: Neutron
pairing energy. Lower: Proton pairing energy.}
\end{figure}

A summary of ground-state properties is shown in
fig.~\ref{fig:obsrefit}.  The binding energies are displayed in terms
of the two-neutron separation energies
${S_{2n}(N,Z)}={E(N-2,Z)-E(N,Z)}$
because this amplifies possible effects. The overall size and trend is
not much affected by the change of the pairing strength. There is,
however, a systematic modification near the shell closures. Consider,
e.g., the magic ${N=82}$. The larger, refitted pairing strength
produces a smaller $S_{2n}$ before this shell and a larger $S_{2n}$
right after it. The jump of the two-neutron separation energies at a
magic shell serves as an empirical measure of the ``magicity'', called
the two-neutron shell gap. And we see that this shell gap is reduced
by as much as $25\%$ when employing the refitted pairing strength. The
binding energies as such are slightly enhanced by at most $0.25\%$ in
the mid-shell region due to the stronger pairing. This change can be,
and needs to be, corrected by a slight readjustment of the RMF
parametrization.

Fig.~\ref{fig:obsrefit} also shows the bulk properties of the nuclear
charge distribution, i.e., root-mean-square radii \cite{Fricke},
diffraction radii, and surface thicknesses \cite{Fri82a}.  There are
small effects visible but they remain far below the precision of the
force to describe these observables (indicated by the error
bars). Density and formfactor are thus rather robust against this
change in pairing strength.

The situation may be different for deformation properties because
these are known to result from an interplay between shell structure
and pairing. As an example, in fig.~\ref{fig:110sn} we show the
quadrupole deformation energy for $^{110}{\rm Sm}$, the softest member
of the Sn isotope chain.  The results confirm the rule-of-thumb that
stronger pairing acts more to restore the spherical shape.  The
refitted, enhanced pairing strength clearly produces a stiffer
deformation energy curve and suppresses more efficiently the deformed
side minima which are visible in case of the standard pairing
strengths. It is worthwhile to estimate the consequences for the
low-lying $2^+$ state. The curvature at ${\beta_2=0}$ is enhanced by
about $130\%$. The cranking mass is proportional to the inverse
quasiparticle energies which are enhanced by about $35\%$ for the
refitted pairing strengths. The mass is thus reduced by $25\%$ and the
energy of the low-lying $2^+$ state is then altogether increased by
about $60\%$ for the refitted pairing strengths.  This little estimate
demonstrates that an appropriate determination of the pairing strength
is crucial for these low-lying collective modes.
\begin{figure}[t!]
\centerline{\epsfig{figure=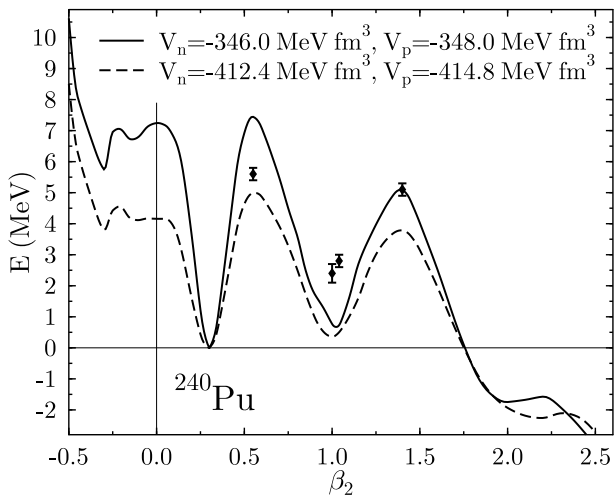,width=8.2cm}}
\caption{\label{fig:240pu} Deformation energy curve  (defined as in
as fig.~\ref{fig:110sn}) for $^{240}{\rm Pu}$.  The points with error
bars show the experimental minima and barriers
\protect\cite{Pubar,TOI}.}
\end{figure}

The two lower plots of fig.~\ref{fig:110sn} show the proton and
neutron pairing energies versus deformation. The larger refitted
strength yields, of course, much larger pairing energies (absolute
values),
but it does also produce a larger diffuseness of occupation numbers
near the Fermi surface which, in turn, degrades the mean-field
binding leaving altogether the net effect of a stiffer deformation
energy curve as seen in the upper panel of the figure. The example
does also show the breakdown of pairing towards the proton shell
closure at spherical shape, and it shows that the breakdown comes, of
course, a bit later for the stronger pairing. This feature, however, 
should not be overstressed because the breakdown is anyway an 
artefact of the mere BCS treatment. A more correct approach would be 
to use particle number projection. But the main effect on the 
deformation energy curve will not be changed by that.

Other important observables in nuclear physics are fission barriers
and it is worth looking at the effect when the pairing strengths are
changed. Fig.~\ref{fig:240pu} shows the energy curve for asymmetric
fission of ${}^{240}{\rm Pu}$.
For this heavy nucleus, we see even more dramatic changes. The
(deformed) minima are softened such that the curvature is reduced by a
factor of two.  At first glance, one may be surprised to see here a
softening whereas the higher pairing strength caused a stronger
curvature in case of ${}^{110}{\rm Sn}$. It is to be noted that the
minima in these two cases have different origins. For ${}^{110}{\rm
Sn}$, the high level density at spherical shape inhibits, in
principle, a spherical minimum and sphericity is only restored by the
action of pairing. Consequently, stronger pairing makes the spherical
shape even more pronounced. For ${}^{240}{\rm Pu}$, on the other hand,
the deformed minimum is caused by a low level density at this place
and pairing can only counteract this preference, thus delivering a
softening of the minimum. There may be, nonetheless, only a small
effect on the energy of the vibrational states in the deformed minima
because the quasiparticle energies (and with them the Inglis mass) are
enhanced also by about a factor of two. On the other hand there is a
strong effect on fission properties because the barriers are lowered
by 1--2 MeV.  The change of the barriers moves them away from the
experimental points for this particular mean-field parametrization
(note that the inner barrier would be further lowered when allowing
for triaxial shapes).  This means that the selection of the
appropriate parametrization for fission \cite{fisspap} needs to be
reconsidered in connection with the new pairing strength.

Altogether, we find that dynamical polarisation effects can have a
strong influence on the pairing gap as deduced from even-odd
staggering of binding energies. In the present test case, one needs to
enhance the underlying pairing strengths by about $20\%$ to compensate
for that effect. This change in pairing strengths has only a small
effect on the ground state properties of even-even nuclei like binding
energies and radii. But more elaborate quantities like the jump of the
two-particle separation energies at magic shells can react
sensitively. Even more dramatic changes are seen for deformation
properties. Vibrations around spherical shapes become more rigid and
fission barriers are significantly lowered. These findings are
disquieting and call for further critical inspection with different
test cases and other mean-field theories. There is, for example, one
possible flaw in the RMF. It is an effective Hartree theory. Thus the
polarisation energy in the odd nucleus contains a contribution from
the self-interaction of the odd nucleon and a self-interaction
corrected theory may produce different results. The situation is
different in the non-relativistic Skyrme--Hartree--Fock model which
can be considered as a true Hartree-Fock variational theory and which
thus is free from the self-interaction effect.
Research in this direction is underway.
%
%============================================================================
%

\bigskip
\noindent
Acknowledgements: We thank W.~Nazarewicz and J.~Dobaczewski for useful
hints and inspiring discussions. This work was supported by 
Bundesministerium f\"ur Bildung und Forschung (BMBF), Project No.~06ER808, 
by Gesellschaft f\"ur Schwerionenforschung (GSI), by the U.S.\ Department 
of Energy under Contract No.~DE-FG02-97ER41019 with the University of 
North Carolina and Contract No.~DE-FG02-96ER40963 with the University 
of Tennessee and by the NATO grant SA.5--2--05 (CRG.971541). The Joint 
Institute for Heavy Ion Research has as member institutions the University 
of Tennessee, Vanderbilt University, and the Oak Ridge National Laboratory; 
it is supported by the members and by the Department of Energy through 
Contract No.~DE-FG05-87ER40361 with the University of Tennessee.
%
%============================================================================
%

\end{document}